\documentclass{nature}
\usepackage{graphicx}
\usepackage{dcolumn}
\usepackage{bm}
\usepackage{color}
\usepackage{lineno}
\usepackage{epsfig}
\usepackage{amsmath}

\title{Extremely large anomalous Hall conductivity and unusual axial diamagnetism in a quasi-1D Dirac material La$_3$MgBi$_5$}

\author{Zhe-Kai Yi,$^{1}$ Peng-Jie Guo,$^{2,\ddag}$ Hui Liang,$^{1}$ Yi-Ran Li,$^{1}$ Ping Su,$^{1}$ Na Li,$^{1}$ Ying Zhou,$^{1}$ Dan-Dan Wu,$^{1}$ Yan Sun,$^{1}$ Xiao-Yu Yue,$^{3}$ Qiu-Ju Li,$^{4}$ Shou-Guo Wang,$^{1}$ Xue-Feng Sun$^{1,5,\S}$ and Yi-Yan Wang$^{1,*}$}

\begin{document}
\bibliographystyle{naturemag}
	\maketitle	
	\begin{affiliations}
		\item Anhui Key Laboratory of Magnetic Functional Materials and Devices, Institute of Physical Science and Information Technology, Anhui University, Hefei, Anhui 230601, China
        \item Department of Physics and Beijing Key Laboratory of Opto-electronic Functional Materials \& Micro-nano Devices, Renmin University of China, Beijing 100872, China
        \item School of Optical and Electronic Information, Suzhou City University, Suzhou, Jiangsu 215104, China
        \item School of Physics and Optoelectronic Engineering, Anhui University, Hefei, Anhui 230601, China
        \item Collaborative Innovation Center of Advanced Microstructures, Nanjing University, Nanjing, Jiangsu 210093, China
	\end{affiliations}
	
	\leftline{$^\ddag$Corresponding authors: guopengjie@ruc.edu.cn}
    \leftline{$^\S$Corresponding authors: xfsun@ahu.edu.cn}
    \leftline{$^*$Corresponding authors: wyy@ahu.edu.cn}\vspace*{1cm}
	
\newpage
\begin{abstract}
Anomalous Hall effect (AHE), one of the most important electronic transport phenomena, generally appears in ferromagnetic materials but is rare in materials without magnetic elements. Here, we present a study of La$_3$MgBi$_5$, whose band structure carries multitype Dirac fermions. Although magnetic elements are absent in La$_3$MgBi$_5$, clear signals of AHE can be observed. In particular, the anomalous Hall conductivity is extremely large, reaching 42,356 $\boldsymbol{\Omega^{-1}}$ cm$^{\boldsymbol{-1}}$ with an anomalous Hall angle of 8.8 \%, the largest one that has been observed in the current AHE systems. The AHE is suggested to originate from the combination of skew scattering and Berry curvature. Another unique property discovered in La$_3$MgBi$_5$ is the axial diamagnetism. The diamagnetism is significantly enhanced and dominates the magnetization in the axial directions, which is the result of restricted motion of the Dirac fermion at Fermi level. Our findings not only establish La$_3$MgBi$_5$ as a suitable platform to study AHE and quantum transport, but also indicate the great potential of 315-type Bi-based materials for exploring novel physical properties.
\end{abstract}

Topological materials provide an ideal platform for exploring various unique physical properties, especially related to the Hall effect, such as anomalous Hall effect (AHE) and quantum anomalous Hall effect\cite{chang2013experimental, deng2020quantum, liu2020robust}. Typically, AHE is mostly observed in ferromagnetic materials with broken time-reversal symmetry. The microscopic origins include extrinsic mechanism related to scattering effects and intrinsic mechanism related to Berry curvature. For example, the ferromagnetic Weyl semimetal Co$_3$Sn$_2$S$_2$ exhibits AHE driven by the large intrinsic Berry curvature\cite{liu2018giant, wang2018large}. The AHE can also appear in antiferromagnetic and non-magnetic materials. Several antiferromagnets have been found to host the AHE\cite{nakatsuji2015large, nayak2016large, suzuki2016large, PhysRevB.107.L161109, PhysRevLett.112.017205, vsmejkal2022anomalous}, including Mn$_3$Sn, Mn$_3$Ge, and GdPtBi etc. ZrTe$_5$ is a famous non-magnetic AHE material\cite{liang2018anomalous}, in which the application of external magnetic field breaks the time-reversal symmetry and leads to the generation of large Berry curvature. Recently, the topological kagome metals KV$_3$Sb$_5$ and CsV$_3$Sb$_5$ have been reported to exhibit AHE and large anomalous Hall conductivity (AHC) in the absence of magnetic ordering\cite{yang2020giant, PhysRevB.104.L041103}. To date, only a few non-magnetic materials can exhibit AHE. It is of interest to explore the AHE and large AHC in topological materials without magnetic elements.

Bismuth, usually exhibiting a valence of -3 in compounds. However, in some special cases, Bi exhibits the hypervalent states, such as the Bi$^{1-}$ in 2D square net and the Bi$^{2-}$ in 1D chain. It is worth noting that the 112-type Mn-based materials with Bi/Sb 2D square net structure have attracted great attention in the development of topological materials\cite{borisenko2019time, liu2017magnetic, liu2017unusual, PhysRevMaterials.2.021201, wang2017magneto}. In the 112-type materials, the Bi/Sb 2D square net structure plays an important role in the formation of topological state. Recently, the interacting 1D Bi chains are theoretically indicated to generate topological electronic structure\cite{khoury2022class}, which motivates us to explore novel physical properties in materials with 1D Bi chains.

The Ln$_3$MPn$_5$ (Ln = lanthanide; M = metal; Pn = pnictide) family is a series of quasi-1D materials and a good carrier for such Bi chains. In this work, we focus on the compound La$_3$MgBi$_5$, a member of the Ln$_3$MPn$_5$ family. We grow the high quality single crystals and study the transport properties and electronic structure. The first-principles calculations show that multitype Dirac fermions coexist in La$_3$MgBi$_5$, including type-I, type-II, higher-order quadratic and nodal-line. Despite the absence of magnetic elements, La$_3$MgBi$_5$ still exhibits a significant AHE. Surprisingly, the obtained AHC is extremely large, reaching 42,356 \emph{$\Omega^{-1}$} cm$^{-1}$ with an anomalous Hall angle of 8.8 \%. The origin of the AHE can be attributed to the combination of skew scattering and Berry curvature. In addition, the angle dependent magnetization measurements reveal the unusual axial diamagnetism in La$_3$MgBi$_5$. In the axial directions, the diamagnetism is significantly enhanced, which comes from the restricted motion of the Dirac fermions at Fermi level.

La$_3$MgBi$_5$, a quasi-1D compound crystallized in the space group \emph{P6}$_3$/\emph{mcm} (No. 193), contains the hypervalent 1D Bi chain in its crystal structure (Figs. 1a,b). As shown in Fig. 1c, the grown crystal is rod-shaped, and the single crystal XRD pattern indicates that the natural surface is (010) plane. The first-principles calculations have been employed to study the electronic structure of La$_3$MgBi$_5$. Figure 1e presents the band structure of La$_3$MgBi$_5$ along high symmetry lines with the spin-orbital coupling (SOC) effect included. Multitype Dirac fermions can be found in the band structure. Along the direction of $\Gamma$-A, there are several symmetry protected Dirac fermions, including the type-I and type-II. In particular, the type-II Dirac point located in the middle of $\Gamma$-A is very close to the Fermi level. At point A, there exists two higher-order quadratic Dirac points with linear dispersion along the principle axis but quadratic dispersion in the plane perpendicular to it\cite{PhysRevX.7.021019, PhysRevB.101.205134}. Along the direction of A-H-T, two electron-type bands overlap, forming the four-fold degenerate Dirac nodal-line.

Due to the presence of multitype Dirac fermions near the Fermi level in La$_3$MgBi$_5$, the exotic physical properties are expected, such as the AHE induced by the large Berry curvature. The electrical transport properties of La$_3$MgBi$_5$ have been studied in detail. Before the measurements, the sample (S4-b) was processed into a rectangular shape with a thickness of 0.06 mm (see Supplementary Information). La$_3$MgBi$_5$ exhibits metallic behavior and large magnetoresistance (331 \% at 1.8 K and 14 T, see Supplementary Information). Figure 2b shows the field dependent Hall resistivity $\rho_{xy}$ of La$_3$MgBi$_5$ at various temperatures. At low temperatures, the $\rho_{xy}$ curves exhibit a tilted N-shaped pattern in low magnetic fields. Obviously, the overall behavior of $\rho_{xy}$ cannot be described using the semiclassical two-band model. To understand the unusual Hall effect in La$_3$MgBi$_5$, we adopt the following formula to fit the data,
\begin{equation}\label{equ1}
\rho_{xy}=\rho_{\rm{AHE}}+\rho_{xy}^{\rm{two\text{-}band}},
\end{equation}
\begin{equation}\label{equ2}
\rho_{\rm{AHE}}=\rho_{\rm{AHE}}^{\enspace 0}tanh(B/B_0),
\end{equation}
\begin{equation}\label{equ3}
\rho_{xy}^{\rm{two\text{-}band}}=\frac{B}{e}\frac{(n_h \mu_h^2-n_e \mu_e^2)+(n_h-n_e)(\mu_h \mu_e)^2 B^2}{(n_h \mu_h+n_e \mu_e)^2+(n_h-n_e)^2 (\mu_h \mu_e)^2 B^2}.
\end{equation}
Here, $\rho_{\rm{AHE}}$ and $\rho_{xy}^{\rm{two\text{-}band}}$ represent the AHE term and the two-band model term, respectively. $\rho_{\rm{AHE}}^{\enspace 0}$ is the saturation value of $\rho_{\rm{AHE}}$. $B_0$ is a parameter related to the saturation field. The hyperbolic tangent function in $\rho_{\rm{AHE}}$ is an empirical approach to describe the behavior of AHE, which is often used in other topological semimetals\cite{PhysRevB.103.L201110, PhysRevLett.118.136601, PhysRevLett.123.196602}. $n_h(n_e)$ and $\mu_h(\mu_e)$ represent the hole (electron) concentration and hole (electron) mobility, respectively. As shown in Fig. 2c, the $\rho_{xy}$ curve can be well fitted by the combination of AHE and two-band model, strongly indicating the existence of AHE in La$_3$MgBi$_5$. Figure 2d shows the separated anomalous component $\rho_{\rm{AHE}}$ after removing the two-band background, which more clearly demonstrates the AHE signals in La$_3$MgBi$_5$. The angle dependence of AHE for the magnetic field rotating in the $bc$ plane and in the plane perpendicular to the $c$ axis has also been investigated. Figures 3b,e and 3c,f show the corresponding Hall resistivity $\rho_{xy}$ and the extracted anomalous Hall resistivity $\rho_{\rm{AHE}}$, respectively. As the magnetic field deviates from the $b$ axis and rotates to the in-plane direction ($\theta$ and $\varphi$ decrease), AHE gradually weakens.

Unexpectedly, we found that the AHC of La$_3$MgBi$_5$ is extremely large, reaching 42,356 \emph{$\Omega^{-1}$} cm$^{-1}$ with an anomalous Hall angle of 8.8 \%. To our knowledge, such a value of AHC is the largest one that has observed in the current AHE systems, even exceeding that of MnGe thin film (160 nm)\cite{fujishiro2021giant}. Especially, the large AHC here appears in non-magnetic bulk material La$_3$MgBi$_5$, making the observed AHE very rare. In various materials\cite{liu2018giant, fujishiro2021giant, PhysRevLett.106.156603, PhysRevLett.99.086602, xu2022topological, zeng2022large, yang2020giant, PhysRevB.104.L041103, suzuki2016large, song2021tunable, PhysRevLett.99.077202, zheng2023electrically}, as shown in Fig. 4, the AHC is usually in the order of 10$^3$ \emph{$\Omega^{-1}$} cm$^{-1}$ or below. The recent kagome materials KV$_3$Sb$_5$ and CsV$_3$Sb$_5$ also exhibited large AHC\cite{yang2020giant, PhysRevB.104.L041103}, reaching 15,507 \emph{$\Omega^{-1}$} cm$^{-1}$ and 2.1$\times$10$^4$ \emph{$\Omega^{-1}$} cm$^{-1}$, respectively.

It is interesting to explore the origin of the extremely large AHC in La$_3$MgBi$_5$. In some non-magnetic or magnetic topological materials, such as ZrTe$_5$\cite{liang2018anomalous}, KZnBi\cite{song2021tunable} and Co$_3$Sn$_2$S$_2$\cite{liu2018giant,wang2018large}, the AHE is identified as induced by the intrinsic Berry curvature. In La$_3$MgBi$_5$ with multitype Dirac fermions, the application of magnetic field breaks the time-reversal symmetry, resulting in the generation of Weyl fermions and Berry curvature. However, the AHC induced by Berry curvature is constrained by the quantization limit $e^2/ha$ ($a$ is the lattice constant)\cite{PhysRevB.77.165103} and usually in the order of 10$^2$$\sim$10$^3$ \emph{$\Omega^{-1}$} cm$^{-1}$. Berry curvature alone is not enough to generate such a large AHC in La$_3$MgBi$_5$. The scaling law between $\sigma_{xy}^{\rm{AHE}}$ and $\sigma_{xx}$ is often employed to identify the possible underlying mechanism of AHE. In the framework of unified model, the AHC driven by Berry curvature has the independence of $\sigma_{xx}$\cite{PhysRevLett.97.126602, PhysRevLett.99.086602, liu2018giant}. In the region of high conductivity, AHE is dominated by the skew scattering mechanism, and $\sigma_{xy}^{\rm{AHE}}$ scales linearly with $\sigma_{xx}$. As shown in Fig. 4, the AHC in La$_3$MgBi$_5$ follows $\sigma_{xy}^{\rm{AHE}}$$\propto$$\sigma_{xx}$, indicating the dominance of skew scattering. Considering the existence of multitype Dirac fermions in La$_3$MgBi$_5$, it is suggested that the observed large AHE originates from the combination of skew scattering and Berry curvature.

We further investigate the generation of skew scattering to gain more insights into the AHE. Skew scattering is an asymmetric scattering process of carriers from nonmagnetic/magnetic impurities. Several mechanisms have been proposed that may lead to asymmetric scattering\cite{RevModPhys.82.1539}. Spin waves of local moments at finite temperature and resonant skew scattering can be excluded for the absence of magnetic ordering in La$_3$MgBi$_5$. Recently, spin-chirality scattering in triangular magnetic lattice was proposed to produce strong skew scattering\cite{PhysRevB.103.235148, fujishiro2021giant}, which is however not the case in La$_3$MgBi$_5$ either. In fact, SOC can cause the asymmetric scattering and introduce a momentum perpendicular to the incident momentum and magnetization, which has been derived in detail in ferromagnetic metal\cite{PhysRevB.77.165103} and 2D massive Dirac band\cite{PhysRevB.75.045315}. Especially, the SOC-related skew scattering can arise from local paramagnetic centers\cite{zheng2023electrically, maryenko2017observation}, even without spin polarization. In La$_3$MgBi$_5$, both paramagnetism and massive Dirac fermions\cite{PhysRevB.108.075157} exist and can contribute to skew scattering.

In addition to the AHE, La$_3$MgBi$_5$ also exhibits unique axial diamagnetism. As shown in Fig. 5a, when the magnetic field is perpendicular to the ${bc}$ plane ($B \perp bc$, this is not in the direction of $a$ axis), the magnetization exhibits strong quantum oscillation and paramagnetism. However, when the magnetic field turns towards the directions of the crystal axis ($B \parallel b$ and $B \parallel c$, Figs. 5b,c), paramagnetism disappears and strong diamagnetism emerges. It should be noted that the $a$ axis and $b$ axis are equivalent in the crystal structure of La$_3$MgBi$_5$. This indicates the axial diamagnetic characteristics of the magnetization. Angle dependent magnetization measurements have been performed to further characterize the magnetic properties of La$_3$MgBi$_5$. As the magnetic field rotates from $B \perp bc$ to $B \parallel c$ (Fig. 5d), paramagnetism gradually weakens and diamagnetism appears. At 14 T, the ratio \emph{M}$_{\psi=90^{\circ}}$/\emph{M}$_{\psi=0^{\circ}}$ is 6.56, indicating a significant increase in diamagnetism in the axial direction of the crystal. In addition, as the magnetic field rotates from $B \parallel c$ to $B \parallel b$ (Fig. 5f), the diamagnetism first weakens and then enhances. The magnetization of $B \parallel b$ is close to that of $B \parallel c$. It can be seen that La$_3$MgBi$_5$ exhibits strong diamagnetism along the axes of the crystal. Once deviated from the axial directions, the diamagnetism will weaken or even be completely suppressed by paramagnetism.

Such an axial diamagnetism is rare in materials. We further explored the underlying mechanism and found that it originates from the Dirac fermion crossing the Fermi level in the A-H direction. For a Dirac type band structure with Dirac point located at the Fermi level\cite{PhysRevB.93.045201}, the spin susceptibility describing Pauli and Van-Vleck paramagnetism is completely Fermi energy independent. The SOC term in the magnetic susceptibility is paramagnetic in the electron side and diamagnetic in the hole side, forming an antisymmetric linear function of the Fermi energy. However, the orbital susceptibility logarithmically diverges in the diamagnetic direction at the Dirac point, and its value is related to Fermi velocity and energy\cite{PhysRevB.93.045201}. In the case of La$_3$MgBi$_5$, there are two Dirac points that may generate significant orbital diamagnetism. One is the type-II Dirac point near the Fermi level in the $\Gamma$-A direction, and the other is the intersection of Dirac nodal-line and Fermi level in the A-H direction. In order to identify the origin of axial diamagnetism, the band structures around the type-II Dirac point have been calculated (Fig. S7, see Supplementary Information). It can be seen that the Dirac point is type-II in the $\Gamma$-A direction, but type-I in the $k_x$-$k_y$ plane. At point D, there is no significant difference or anisotropy in the band structure along the B-D-B and C-D-C directions, indicating that this Dirac point is unlikely to cause magnetization difference in $B \perp bc$ (paramagnetism) and $B \parallel b$ (diamagnetism). Now we turn to the intersection of Dirac nodal-line and Fermi level in the A-H direction. Compared to the type-II Dirac point in the $\Gamma$-A direction, this Dirac point crosses the Fermi level and has a larger number. As shown in Fig. 5e, this Dirac point appears in the corresponding vertical planes for $B \parallel b$ and $B \parallel c$, but is absent for $B \perp bc$. As a result, when the magnetic field is applied along the crystal axes, the orbital diamagnetism of the Dirac fermion begins to take effect and dominate. When the magnetic field deviates from the axial directions, the orbital diamagnetism of the Dirac fermion weakens and is suppressed by the paramagnetism of other bands. The axial diamagnetism is the result of restricted motion of the Dirac fermion at Fermi level, which may also be used to explain the highly anisotropic magnetization in other topological materials.

In summary, we report the observation of extremely large AHC and axial diamagnetism in the single crystal of La$_3$MgBi$_5$, which hosts the multitype Dirac fermions. Even if there are no magnetic elements in La$_3$MgBi$_5$, obvious AHE signals can be observed with an extremely large AHC value of 42,356 \emph{$\Omega^{-1}$} cm$^{-1}$. This large AHE comes from the combination of skew scattering and Berry curvature. La$_3$MgBi$_5$ also exhibits unusual axial diamagnetism. The diamagnetism is significantly strengthened and dominates in the axial directions, which can be attributed to the restricted motion of the Dirac fermion at Fermi level.

\emph{Note added in proof:} As this article was about to be completed, we noticed another related work on La$_3$MgBi$_5$, which reported the quantum oscillations and nontrivial band topology in La$_3$MgBi$_5$\cite{PhysRevB.108.075157}.

\section*{Methods}
\subsection{Crystal growth and transport measurements}
Large single crystals of La$_3$MgBi$_5$ were grown by the flux method. The starting reactants, including La (powder), Mg (granules), and Bi (granules), were mixed in a molar ratio of La: Mg: Bi = 1: 3: 6.5 and placed into an alumina crucible. The operation was carried out in an Ar-filled glove box. The crucible was then sealed into an evacuated quartz tube and heated to 850 $^\circ$C. After maintaining this temperature for 20 h, the temperature was slowly reduced to 650 $^\circ$C at a rate of 1 $^\circ$C/h. Then the excess flux was separated from the crystals through a centrifuge. The obtained single crystals of La$_3$MgBi$_5$ exhibit rod-shaped morphology. The atomic proportion has been checked by energy dispersive X-ray spectroscopy (EDS). The X-ray diffraction (XRD) pattern of single crystal was collected from a SmartLab X-ray diffractometer. The measurements of transport properties were performed on a Quantum Design physical property measurement system (PPMS). The magnetic properties were measured with a vibrating sample magnetometer (VSM) option.

\subsection{Electronic structure calculations}
The first-principles electronic structure calculations were performed with the projector augmented wave (PAW) method\cite{PhysRevB.50.17953, PhysRevB.59.1758} as implemented in the Vienna ab initio simulation package (VASP)\cite{PhysRevB.47.558, kresse1996efficiency, PhysRevB.54.11169}. The generalized gradient approximation (GGA) of Perdew-Burke-Ernerh (PBE) type\cite{PhysRevLett.77.3865} was adopted for the exchange-correlation potential. The kinetic energy cutoff of the plane wave basis was set to be 400 eV. For the Brillouin zone sampling, a 8$\times$8$\times$12 $k$-point mesh was employed. The Gaussian smearing with a width of 0.01 eV was used around the Fermi surface. In structural optimization, both cell parameters and internal atomic positions were allowed to relax until all forces on atoms were smaller than 0.01 eV/{\AA}. The relaxed structure parameters are well agreed with experimental results. The SOC effect was included in the calculations of the electronic properties. The Fermi surfaces are calculated by wannier90 package\cite{PhysRevB.56.12847, PhysRevB.65.035109}.

\section*{Data availability}
The authors declare that the data supporting the findings of this study are available within the article and its Supplementary Information. Extra data are available from the corresponding authors upon reasonable request.

\section*{References}
\bibliography{Bibtex}

\begin{thebibliography}{10}
\expandafter\ifx\csname url\endcsname\relax
  \def\url#1{\texttt{#1}}\fi
\expandafter\ifx\csname urlprefix\endcsname\relax\def\urlprefix{URL }\fi
\providecommand{\bibinfo}[2]{#2}
\providecommand{\eprint}[2][]{\url{#2}}

\bibitem{chang2013experimental}
\bibinfo{author}{Chang, C.-Z.} \emph{et~al.}
\newblock \bibinfo{title}{{Experimental observation of the quantum anomalous
  Hall effect in a magnetic topological insulator}}.
\newblock \emph{\bibinfo{journal}{Science}} \textbf{\bibinfo{volume}{340}},
  \bibinfo{pages}{167--170} (\bibinfo{year}{2013}).

\bibitem{deng2020quantum}
\bibinfo{author}{Deng, Y.} \emph{et~al.}
\newblock \bibinfo{title}{{Quantum anomalous Hall effect in intrinsic magnetic
  topological insulator MnBi$_2$Te$_4$}}.
\newblock \emph{\bibinfo{journal}{Science}} \textbf{\bibinfo{volume}{367}},
  \bibinfo{pages}{895--900} (\bibinfo{year}{2020}).

\bibitem{liu2020robust}
\bibinfo{author}{Liu, C.} \emph{et~al.}
\newblock \bibinfo{title}{{Robust axion insulator and Chern insulator phases in
  a two-dimensional antiferromagnetic topological insulator}}.
\newblock \emph{\bibinfo{journal}{Nat. Mater.}} \textbf{\bibinfo{volume}{19}},
  \bibinfo{pages}{522--527} (\bibinfo{year}{2020}).

\bibitem{liu2018giant}
\bibinfo{author}{Liu, E.} \emph{et~al.}
\newblock \bibinfo{title}{{Giant anomalous Hall effect in a ferromagnetic
  kagome-lattice semimetal}}.
\newblock \emph{\bibinfo{journal}{Nat. Phys.}} \textbf{\bibinfo{volume}{14}},
  \bibinfo{pages}{1125--1131} (\bibinfo{year}{2018}).

\bibitem{wang2018large}
\bibinfo{author}{Wang, Q.} \emph{et~al.}
\newblock \bibinfo{title}{{Large intrinsic anomalous Hall effect in
  half-metallic ferromagnet Co$_3$Sn$_2$S$_2$ with magnetic Weyl fermions}}.
\newblock \emph{\bibinfo{journal}{Nat. Commun.}} \textbf{\bibinfo{volume}{9}},
  \bibinfo{pages}{3681} (\bibinfo{year}{2018}).

\bibitem{nakatsuji2015large}
\bibinfo{author}{Nakatsuji, S.}, \bibinfo{author}{Kiyohara, N.} \&
  \bibinfo{author}{Higo, T.}
\newblock \bibinfo{title}{{Large anomalous Hall effect in a non-collinear
  antiferromagnet at room temperature}}.
\newblock \emph{\bibinfo{journal}{Nature}} \textbf{\bibinfo{volume}{527}},
  \bibinfo{pages}{212--215} (\bibinfo{year}{2015}).

\bibitem{nayak2016large}
\bibinfo{author}{Nayak, A.~K.} \emph{et~al.}
\newblock \bibinfo{title}{{Large anomalous Hall effect driven by a nonvanishing
  Berry curvature in the noncolinear antiferromagnet Mn$_3$Ge}}.
\newblock \emph{\bibinfo{journal}{Sci. Adv.}} \textbf{\bibinfo{volume}{2}},
  \bibinfo{pages}{e1501870} (\bibinfo{year}{2016}).

\bibitem{suzuki2016large}
\bibinfo{author}{Suzuki, T.} \emph{et~al.}
\newblock \bibinfo{title}{{Large anomalous Hall effect in a half-Heusler
  antiferromagnet}}.
\newblock \emph{\bibinfo{journal}{Nat. Phys.}} \textbf{\bibinfo{volume}{12}},
  \bibinfo{pages}{1119--1123} (\bibinfo{year}{2016}).

\bibitem{PhysRevB.107.L161109}
\bibinfo{author}{Hou, X.-Y.}, \bibinfo{author}{Yang, H.-C.},
  \bibinfo{author}{Liu, Z.-X.}, \bibinfo{author}{Guo, P.-J.} \&
  \bibinfo{author}{Lu, Z.-Y.}
\newblock \bibinfo{title}{{Large intrinsic anomalous Hall effect in both
  Nb$_2$FeB$_2$ and Ta$_2$FeB$_2$ with collinear antiferromagnetism}}.
\newblock \emph{\bibinfo{journal}{Phys. Rev. B}}
  \textbf{\bibinfo{volume}{107}}, \bibinfo{pages}{L161109}
  (\bibinfo{year}{2023}).

\bibitem{PhysRevLett.112.017205}
\bibinfo{author}{Chen, H.}, \bibinfo{author}{Niu, Q.} \&
  \bibinfo{author}{MacDonald, A.~H.}
\newblock \bibinfo{title}{{Anomalous Hall Effect Arising from Noncollinear
  Antiferromagnetism}}.
\newblock \emph{\bibinfo{journal}{Phys. Rev. Lett.}}
  \textbf{\bibinfo{volume}{112}}, \bibinfo{pages}{017205}
  (\bibinfo{year}{2014}).

\bibitem{vsmejkal2022anomalous}
\bibinfo{author}{{\v{S}}mejkal, L.}, \bibinfo{author}{MacDonald, A.~H.},
  \bibinfo{author}{Sinova, J.}, \bibinfo{author}{Nakatsuji, S.} \&
  \bibinfo{author}{Jungwirth, T.}
\newblock \bibinfo{title}{{Anomalous Hall antiferromagnets}}.
\newblock \emph{\bibinfo{journal}{Nat. Rev. Mater.}}
  \textbf{\bibinfo{volume}{7}}, \bibinfo{pages}{482--496}
  (\bibinfo{year}{2022}).

\bibitem{liang2018anomalous}
\bibinfo{author}{Liang, T.} \emph{et~al.}
\newblock \bibinfo{title}{{Anomalous Hall effect in ZrTe$_5$}}.
\newblock \emph{\bibinfo{journal}{Nat. Phys.}} \textbf{\bibinfo{volume}{14}},
  \bibinfo{pages}{451--455} (\bibinfo{year}{2018}).

\bibitem{yang2020giant}
\bibinfo{author}{Yang, S.-Y.} \emph{et~al.}
\newblock \bibinfo{title}{{Giant, unconventional anomalous Hall effect in the
  metallic frustrated magnet candidate, KV$_3$Sb$_5$}}.
\newblock \emph{\bibinfo{journal}{Sci. Adv.}} \textbf{\bibinfo{volume}{6}},
  \bibinfo{pages}{eabb6003} (\bibinfo{year}{2020}).

\bibitem{PhysRevB.104.L041103}
\bibinfo{author}{Yu, F.~H.} \emph{et~al.}
\newblock \bibinfo{title}{{Concurrence of anomalous Hall effect and charge
  density wave in a superconducting topological kagome metal}}.
\newblock \emph{\bibinfo{journal}{Phys. Rev. B}}
  \textbf{\bibinfo{volume}{104}}, \bibinfo{pages}{L041103}
  (\bibinfo{year}{2021}).

\bibitem{borisenko2019time}
\bibinfo{author}{Borisenko, S.} \emph{et~al.}
\newblock \bibinfo{title}{{Time-reversal symmetry breaking type-II Weyl state
  in YbMnBi$_2$}}.
\newblock \emph{\bibinfo{journal}{Nat. Commun.}} \textbf{\bibinfo{volume}{10}},
  \bibinfo{pages}{3424} (\bibinfo{year}{2019}).

\bibitem{liu2017magnetic}
\bibinfo{author}{Liu, J.} \emph{et~al.}
\newblock \bibinfo{title}{{A magnetic topological semimetal
  Sr$_{1-y}$Mn$_{1-z}$Sb$_2$ (y, z $<$ 0.1)}}.
\newblock \emph{\bibinfo{journal}{Nat. Mater.}} \textbf{\bibinfo{volume}{16}},
  \bibinfo{pages}{905--910} (\bibinfo{year}{2017}).

\bibitem{liu2017unusual}
\bibinfo{author}{Liu, J.} \emph{et~al.}
\newblock \bibinfo{title}{{Unusual interlayer quantum transport behavior caused
  by the zeroth Landau level in YbMnBi$_2$}}.
\newblock \emph{\bibinfo{journal}{Nat. Commun.}} \textbf{\bibinfo{volume}{8}},
  \bibinfo{pages}{646} (\bibinfo{year}{2017}).

\bibitem{PhysRevMaterials.2.021201}
\bibinfo{author}{Wang, Y.-Y.}, \bibinfo{author}{Xu, S.}, \bibinfo{author}{Sun,
  L.-L.} \& \bibinfo{author}{Xia, T.-L.}
\newblock \bibinfo{title}{{Quantum oscillations and coherent interlayer
  transport in a new topological Dirac semimetal candidate YbMnSb$_2$}}.
\newblock \emph{\bibinfo{journal}{Phys. Rev. Mater.}}
  \textbf{\bibinfo{volume}{2}}, \bibinfo{pages}{021201} (\bibinfo{year}{2018}).

\bibitem{wang2017magneto}
\bibinfo{author}{Wang, Y.-Y.} \emph{et~al.}
\newblock \bibinfo{title}{{Magneto-transport and electronic structures of
  BaZnBi$_2$}}.
\newblock \emph{\bibinfo{journal}{New J. Phys.}} \textbf{\bibinfo{volume}{19}},
  \bibinfo{pages}{123044} (\bibinfo{year}{2017}).

\bibitem{khoury2022class}
\bibinfo{author}{Khoury, J.~F.} \emph{et~al.}
\newblock \bibinfo{title}{{A Class of Magnetic Topological Material Candidates
  with Hypervalent Bi Chains}}.
\newblock \emph{\bibinfo{journal}{J. Am. Chem. Soc.}}
  \textbf{\bibinfo{volume}{144}}, \bibinfo{pages}{9785--9796}
  (\bibinfo{year}{2022}).

\bibitem{PhysRevX.7.021019}
\bibinfo{author}{Liu, Q.} \& \bibinfo{author}{Zunger, A.}
\newblock \bibinfo{title}{{Predicted Realization of Cubic Dirac Fermion in
  Quasi-One-Dimensional Transition-Metal Monochalcogenides}}.
\newblock \emph{\bibinfo{journal}{Phys. Rev. X}} \textbf{\bibinfo{volume}{7}},
  \bibinfo{pages}{021019} (\bibinfo{year}{2017}).

\bibitem{PhysRevB.101.205134}
\bibinfo{author}{Wu, W.}, \bibinfo{author}{Yu, Z.-M.}, \bibinfo{author}{Zhou,
  X.}, \bibinfo{author}{Zhao, Y.~X.} \& \bibinfo{author}{Yang, S.~A.}
\newblock \bibinfo{title}{{Higher-order Dirac fermions in three dimensions}}.
\newblock \emph{\bibinfo{journal}{Phys. Rev. B}}
  \textbf{\bibinfo{volume}{101}}, \bibinfo{pages}{205134}
  (\bibinfo{year}{2020}).

\bibitem{PhysRevB.103.L201110}
\bibinfo{author}{Liu, Y.} \emph{et~al.}
\newblock \bibinfo{title}{{Induced anomalous Hall effect of massive Dirac
  fermions in ZrTe$_5$ and HfTe$_5$ thin flakes}}.
\newblock \emph{\bibinfo{journal}{Phys. Rev. B}}
  \textbf{\bibinfo{volume}{103}}, \bibinfo{pages}{L201110}
  (\bibinfo{year}{2021}).

\bibitem{PhysRevLett.118.136601}
\bibinfo{author}{Liang, T.} \emph{et~al.}
\newblock \bibinfo{title}{{Anomalous Nernst Effect in the Dirac Semimetal
  Cd$_3$As$_2$}}.
\newblock \emph{\bibinfo{journal}{Phys. Rev. Lett.}}
  \textbf{\bibinfo{volume}{118}}, \bibinfo{pages}{136601}
  (\bibinfo{year}{2017}).

\bibitem{PhysRevLett.123.196602}
\bibinfo{author}{Zhang, J.~L.} \emph{et~al.}
\newblock \bibinfo{title}{{Anomalous Thermoelectric Effects of ZrTe$_5$ in and
  beyond the Quantum Limit}}.
\newblock \emph{\bibinfo{journal}{Phys. Rev. Lett.}}
  \textbf{\bibinfo{volume}{123}}, \bibinfo{pages}{196602}
  (\bibinfo{year}{2019}).

\bibitem{fujishiro2021giant}
\bibinfo{author}{Fujishiro, Y.} \emph{et~al.}
\newblock \bibinfo{title}{{Giant anomalous Hall effect from spin-chirality
  scattering in a chiral magnet}}.
\newblock \emph{\bibinfo{journal}{Nat. Commun.}} \textbf{\bibinfo{volume}{12}},
  \bibinfo{pages}{317} (\bibinfo{year}{2021}).

\bibitem{PhysRevLett.106.156603}
\bibinfo{author}{Kanazawa, N.} \emph{et~al.}
\newblock \bibinfo{title}{{Large Topological Hall Effect in a Short-Period
  Helimagnet MnGe}}.
\newblock \emph{\bibinfo{journal}{Phys. Rev. Lett.}}
  \textbf{\bibinfo{volume}{106}}, \bibinfo{pages}{156603}
  (\bibinfo{year}{2011}).

\bibitem{PhysRevLett.99.086602}
\bibinfo{author}{Miyasato, T.} \emph{et~al.}
\newblock \bibinfo{title}{{Crossover Behavior of the Anomalous Hall Effect and
  Anomalous Nernst Effect in Itinerant Ferromagnets}}.
\newblock \emph{\bibinfo{journal}{Phys. Rev. Lett.}}
  \textbf{\bibinfo{volume}{99}}, \bibinfo{pages}{086602}
  (\bibinfo{year}{2007}).

\bibitem{xu2022topological}
\bibinfo{author}{Xu, X.} \emph{et~al.}
\newblock \bibinfo{title}{{Topological charge-entropy scaling in kagome Chern
  magnet TbMn$_6$Sn$_6$}}.
\newblock \emph{\bibinfo{journal}{Nat. Commun.}} \textbf{\bibinfo{volume}{13}},
  \bibinfo{pages}{1197} (\bibinfo{year}{2022}).

\bibitem{zeng2022large}
\bibinfo{author}{Zeng, H.} \emph{et~al.}
\newblock \bibinfo{title}{{Large anomalous Hall effect in kagom{\'e}
  ferrimagnetic HoMn$_6$Sn$_6$ single crystal}}.
\newblock \emph{\bibinfo{journal}{J. Alloy. Compd.}}
  \textbf{\bibinfo{volume}{899}}, \bibinfo{pages}{163356}
  (\bibinfo{year}{2022}).

\bibitem{song2021tunable}
\bibinfo{author}{Song, J.} \emph{et~al.}
\newblock \bibinfo{title}{{Tunable Berry curvature and transport crossover in
  topological Dirac semimetal KZnBi}}.
\newblock \emph{\bibinfo{journal}{npj Quantum Mater.}}
  \textbf{\bibinfo{volume}{6}}, \bibinfo{pages}{77} (\bibinfo{year}{2021}).

\bibitem{PhysRevLett.99.077202}
\bibinfo{author}{Iguchi, S.}, \bibinfo{author}{Hanasaki, N.} \&
  \bibinfo{author}{Tokura, Y.}
\newblock \bibinfo{title}{{Scaling of Anomalous Hall Resistivity in
  Nd$_2$(Mo$_{1-x}$Nb$_x$)$_2$O$_7$ with Spin Chirality}}.
\newblock \emph{\bibinfo{journal}{Phys. Rev. Lett.}}
  \textbf{\bibinfo{volume}{99}}, \bibinfo{pages}{077202}
  (\bibinfo{year}{2007}).

\bibitem{zheng2023electrically}
\bibinfo{author}{Zheng, G.} \emph{et~al.}
\newblock \bibinfo{title}{{Electrically controlled superconductor-to-failed
  insulator transition and giant anomalous Hall effect in kagome metal
  CsV$_3$Sb$_5$ nanoflakes}}.
\newblock \emph{\bibinfo{journal}{Nat. Commun.}} \textbf{\bibinfo{volume}{14}},
  \bibinfo{pages}{678} (\bibinfo{year}{2023}).

\bibitem{PhysRevB.77.165103}
\bibinfo{author}{Onoda, S.}, \bibinfo{author}{Sugimoto, N.} \&
  \bibinfo{author}{Nagaosa, N.}
\newblock \bibinfo{title}{{Quantum transport theory of anomalous electric,
  thermoelectric, and thermal Hall effects in ferromagnets}}.
\newblock \emph{\bibinfo{journal}{Phys. Rev. B}} \textbf{\bibinfo{volume}{77}},
  \bibinfo{pages}{165103} (\bibinfo{year}{2008}).

\bibitem{PhysRevLett.97.126602}
\bibinfo{author}{Onoda, S.}, \bibinfo{author}{Sugimoto, N.} \&
  \bibinfo{author}{Nagaosa, N.}
\newblock \bibinfo{title}{{Intrinsic Versus Extrinsic Anomalous Hall Effect in
  Ferromagnets}}.
\newblock \emph{\bibinfo{journal}{Phys. Rev. Lett.}}
  \textbf{\bibinfo{volume}{97}}, \bibinfo{pages}{126602}
  (\bibinfo{year}{2006}).

\bibitem{RevModPhys.82.1539}
\bibinfo{author}{Nagaosa, N.}, \bibinfo{author}{Sinova, J.},
  \bibinfo{author}{Onoda, S.}, \bibinfo{author}{MacDonald, A.~H.} \&
  \bibinfo{author}{Ong, N.~P.}
\newblock \bibinfo{title}{{Anomalous Hall effect}}.
\newblock \emph{\bibinfo{journal}{Rev. Mod. Phys.}}
  \textbf{\bibinfo{volume}{82}}, \bibinfo{pages}{1539--1592}
  (\bibinfo{year}{2010}).

\bibitem{PhysRevB.103.235148}
\bibinfo{author}{Ishizuka, H.} \& \bibinfo{author}{Nagaosa, N.}
\newblock \bibinfo{title}{{Large anomalous Hall effect and spin Hall effect by
  spin-cluster scattering in the strong-coupling limit}}.
\newblock \emph{\bibinfo{journal}{Phys. Rev. B}}
  \textbf{\bibinfo{volume}{103}}, \bibinfo{pages}{235148}
  (\bibinfo{year}{2021}).

\bibitem{PhysRevB.75.045315}
\bibinfo{author}{Sinitsyn, N.~A.}, \bibinfo{author}{MacDonald, A.~H.},
  \bibinfo{author}{Jungwirth, T.}, \bibinfo{author}{Dugaev, V.~K.} \&
  \bibinfo{author}{Sinova, J.}
\newblock \bibinfo{title}{{Anomalous Hall effect in a two-dimensional Dirac
  band: The link between the Kubo-Streda formula and the semiclassical
  Boltzmann equation approach}}.
\newblock \emph{\bibinfo{journal}{Phys. Rev. B}} \textbf{\bibinfo{volume}{75}},
  \bibinfo{pages}{045315} (\bibinfo{year}{2007}).

\bibitem{maryenko2017observation}
\bibinfo{author}{Maryenko, D.} \emph{et~al.}
\newblock \bibinfo{title}{{Observation of anomalous Hall effect in a
  non-magnetic two-dimensional electron system}}.
\newblock \emph{\bibinfo{journal}{Nat. Commun.}} \textbf{\bibinfo{volume}{8}},
  \bibinfo{pages}{14777} (\bibinfo{year}{2017}).

\bibitem{PhysRevB.108.075157}
\bibinfo{author}{Han, X.} \emph{et~al.}
\newblock \bibinfo{title}{{Quantum oscillations and transport evidence of
  topological bands in La$_3$MgBi$_5$ single crystals}}.
\newblock \emph{\bibinfo{journal}{Phys. Rev. B}}
  \textbf{\bibinfo{volume}{108}}, \bibinfo{pages}{075157}
  (\bibinfo{year}{2023}).

\bibitem{PhysRevB.93.045201}
\bibinfo{author}{Koshino, M.} \& \bibinfo{author}{Hizbullah, I.~F.}
\newblock \bibinfo{title}{{Magnetic susceptibility in three-dimensional nodal
  semimetals}}.
\newblock \emph{\bibinfo{journal}{Phys. Rev. B}} \textbf{\bibinfo{volume}{93}},
  \bibinfo{pages}{045201} (\bibinfo{year}{2016}).

\bibitem{PhysRevB.50.17953}
\bibinfo{author}{Bl{\"o}chl, P.~E.}
\newblock \bibinfo{title}{Projector augmented-wave method}.
\newblock \emph{\bibinfo{journal}{Phys. Rev. B}} \textbf{\bibinfo{volume}{50}},
  \bibinfo{pages}{17953--17979} (\bibinfo{year}{1994}).

\bibitem{PhysRevB.59.1758}
\bibinfo{author}{Kresse, G.} \& \bibinfo{author}{Joubert, D.}
\newblock \bibinfo{title}{From ultrasoft pseudopotentials to the projector
  augmented-wave method}.
\newblock \emph{\bibinfo{journal}{Phys. Rev. B}} \textbf{\bibinfo{volume}{59}},
  \bibinfo{pages}{1758--1775} (\bibinfo{year}{1999}).

\bibitem{PhysRevB.47.558}
\bibinfo{author}{Kresse, G.} \& \bibinfo{author}{Hafner, J.}
\newblock \bibinfo{title}{Ab initio molecular dynamics for liquid metals}.
\newblock \emph{\bibinfo{journal}{Phys. Rev. B}} \textbf{\bibinfo{volume}{47}},
  \bibinfo{pages}{558--561} (\bibinfo{year}{1993}).

\bibitem{kresse1996efficiency}
\bibinfo{author}{Kresse, G.} \& \bibinfo{author}{Furthm{\"u}ller, J.}
\newblock \bibinfo{title}{Efficiency of ab-initio total energy calculations for
  metals and semiconductors using a plane-wave basis set}.
\newblock \emph{\bibinfo{journal}{Comp. Mater. Sci.}}
  \textbf{\bibinfo{volume}{6}}, \bibinfo{pages}{15--50} (\bibinfo{year}{1996}).

\bibitem{PhysRevB.54.11169}
\bibinfo{author}{Kresse, G.} \& \bibinfo{author}{Furthm{\"u}ller, J.}
\newblock \bibinfo{title}{Efficient iterative schemes for ab initio
  total-energy calculations using a plane-wave basis set}.
\newblock \emph{\bibinfo{journal}{Phys. Rev. B}} \textbf{\bibinfo{volume}{54}},
  \bibinfo{pages}{11169--11186} (\bibinfo{year}{1996}).

\bibitem{PhysRevLett.77.3865}
\bibinfo{author}{Perdew, J.~P.}, \bibinfo{author}{Burke, K.} \&
  \bibinfo{author}{Ernzerhof, M.}
\newblock \bibinfo{title}{Generalized gradient approximation made simple}.
\newblock \emph{\bibinfo{journal}{Phys. Rev. Lett.}}
  \textbf{\bibinfo{volume}{77}}, \bibinfo{pages}{3865--3868}
  (\bibinfo{year}{1996}).

\bibitem{PhysRevB.56.12847}
\bibinfo{author}{Marzari, N.} \& \bibinfo{author}{Vanderbilt, D.}
\newblock \bibinfo{title}{{Maximally localized generalized Wannier functions
  for composite energy bands}}.
\newblock \emph{\bibinfo{journal}{Phys. Rev. B}} \textbf{\bibinfo{volume}{56}},
  \bibinfo{pages}{12847--12865} (\bibinfo{year}{1997}).

\bibitem{PhysRevB.65.035109}
\bibinfo{author}{Souza, I.}, \bibinfo{author}{Marzari, N.} \&
  \bibinfo{author}{Vanderbilt, D.}
\newblock \bibinfo{title}{{Maximally localized Wannier functions for entangled
  energy bands}}.
\newblock \emph{\bibinfo{journal}{Phys. Rev. B}} \textbf{\bibinfo{volume}{65}},
  \bibinfo{pages}{035109} (\bibinfo{year}{2001}).

\end{thebibliography}

\begin{addendum}
\item This work is supported by the National Natural Science Foundation of China (Nos. 12104011, 12274388, 12204533, 52102333, 12104010, 12204004, and 12004003) and the Natural Science Foundation of Anhui Province (Nos. 2108085QA22 and 2108085MA16).

\item[Author contributions] Y.-Y.W. coordinated the project and designed the experiments. Z.-K.Y. and Y.-Y.W. synthesized the single crystals of La$_3$MgBi$_5$. P.-J.G. performed \emph{ab initio} calculations. Y.-Y.W. performed the transport measurements with the assistance of Z.-K.Y.. Z.-K.Y., Y.-Y.W. and P.-J.G. plotted the figures and analysed the experimental data. Y.-Y.W., Z.-K.Y. and X.-F.S. wrote the paper. All authors discussed the results and commented on the manuscript.

\item[Supplementary Information] accompanies this paper.

\item[Author Information] The authors declare no competing interests. The data that support the findings of this study are available from the corresponding authors P.-J.G. (guopengjie@ruc.edu.cn), X.-F.S. (xfsun@ahu.edu.cn) and Y.-Y.W. (wyy@ahu.edu.cn) upon reasonable request.

\end{addendum}
\newpage

\begin{figure*}[htbp]
	\centerline{\epsfig{figure=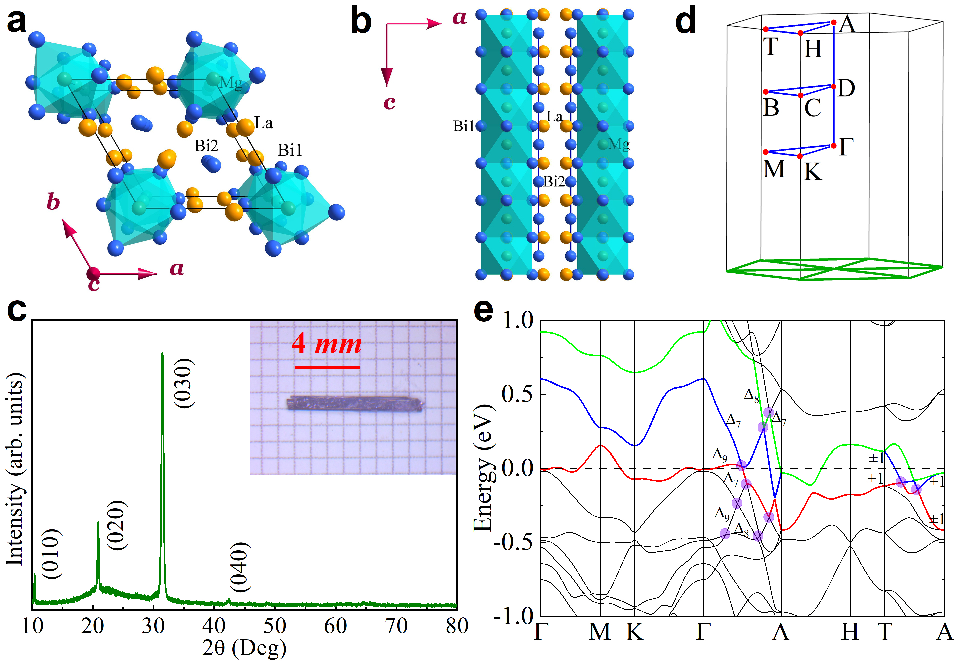,width=0.9\columnwidth}}
	\caption{\textbf{Crystal structure and electronic structure of La$_3$MgBi$_5$.} \textbf{a} and \textbf{b}, Views of the structure from the directions of $c$ axis and $b$ axis, respectively. \textbf{c}, Single crystal XRD pattern of La$_3$MgBi$_5$. Inset: A typical image of the grown crystal. \textbf{d}, Schematic of the 3D Brillouin zone. The dark green lines indicate the Dirac nodal-line along the A-H-T direction. D indicates the position of the type-II Dirac point along the $\Gamma$-A direction. \textbf{e}, The calculated band structure of La$_3$MgBi$_5$ along high symmetry lines with the SOC effect included. The circles indicate the identified type-I and type-II Dirac points.}
\end{figure*}

\begin{figure*}
	\centerline{\epsfig{figure=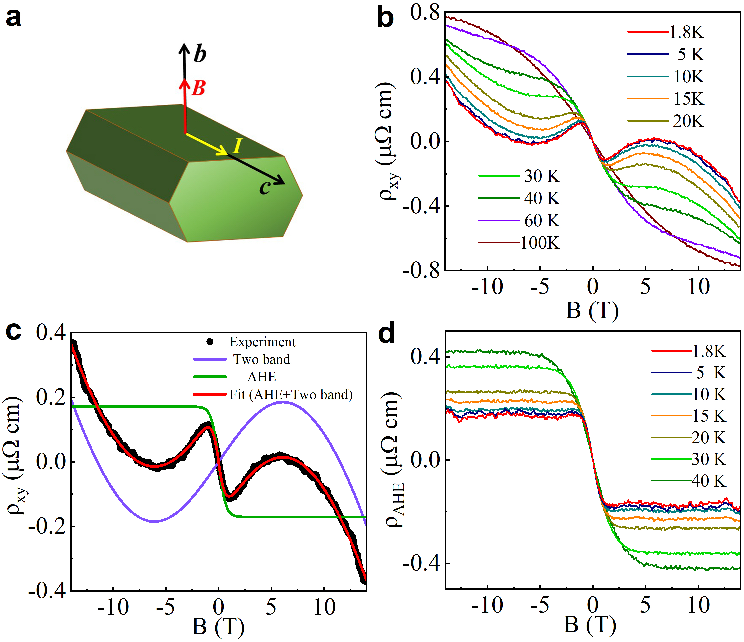,width=0.7\columnwidth}}
	\caption{\textbf{Anomalous Hall effect in La$_3$MgBi$_5$ (Sample 4-b).} \textbf{a}, Schematic of the measurements. The current and magnetic field follow the $c$-axis and $b$-axis directions, respectively. The sample 4-b has been processed into a rectangular shape with a thickness of 0.06 mm. \textbf{b}, The field dependent Hall resistivity at different temperatures. \textbf{c}, Analysis and fitting of the Hall resistivity at 1.8 K. The black dots are experimental data. The solid red line represents the fitting using a combination of AHE and two-band model. The solid green and violet lines represent the AHE curve and the two-band curve separated from the fitting. \textbf{d}, The anomalous Hall resistivity $\rho_{\rm{AHE}}$ extracted from \textbf{b} by removing the two-band background.}
\end{figure*}

\begin{figure*}
	\centerline{\epsfig{figure=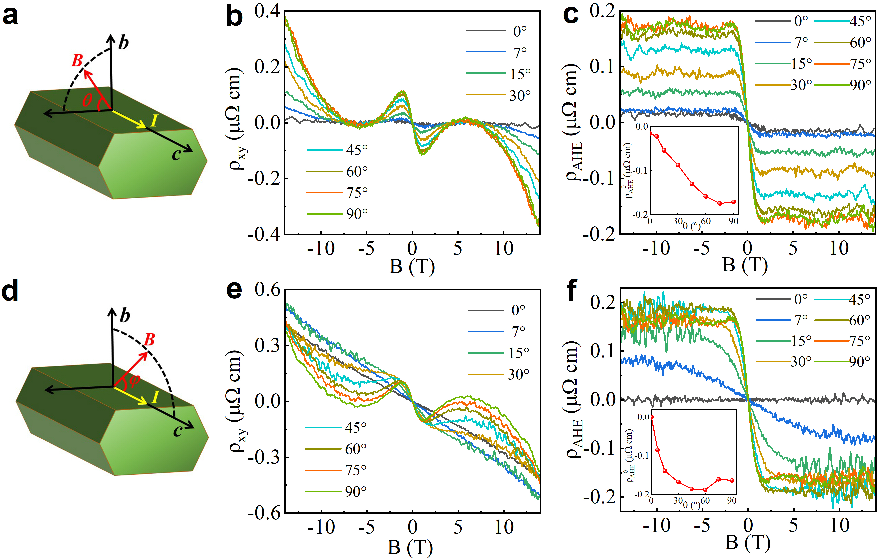,width=0.9\columnwidth}}
	\caption{\textbf{Angle dependence of the AHE.} \textbf{a} and \textbf{d}, Schematic of two configurations of the measurements. The definitions of $\theta$ and $\varphi$ are given in the figures. \textbf{b} and \textbf{e}, The field dependent Hall resistivity at different angles ($\theta$ and $\varphi$). The temperature is fixed at 1.8 K. \textbf{c} and \textbf{f}, The anomalous Hall resistivity $\rho_{\rm{AHE}}$ extracted from \textbf{b} and \textbf{e} by removing the two-band curves. Insets: Angle dependent saturation value ($\rho_{\rm{AHE}}^{\enspace 0}$) of $\rho_{\rm{AHE}}$.}
\end{figure*}

\begin{figure*}
	\centerline{\epsfig{figure=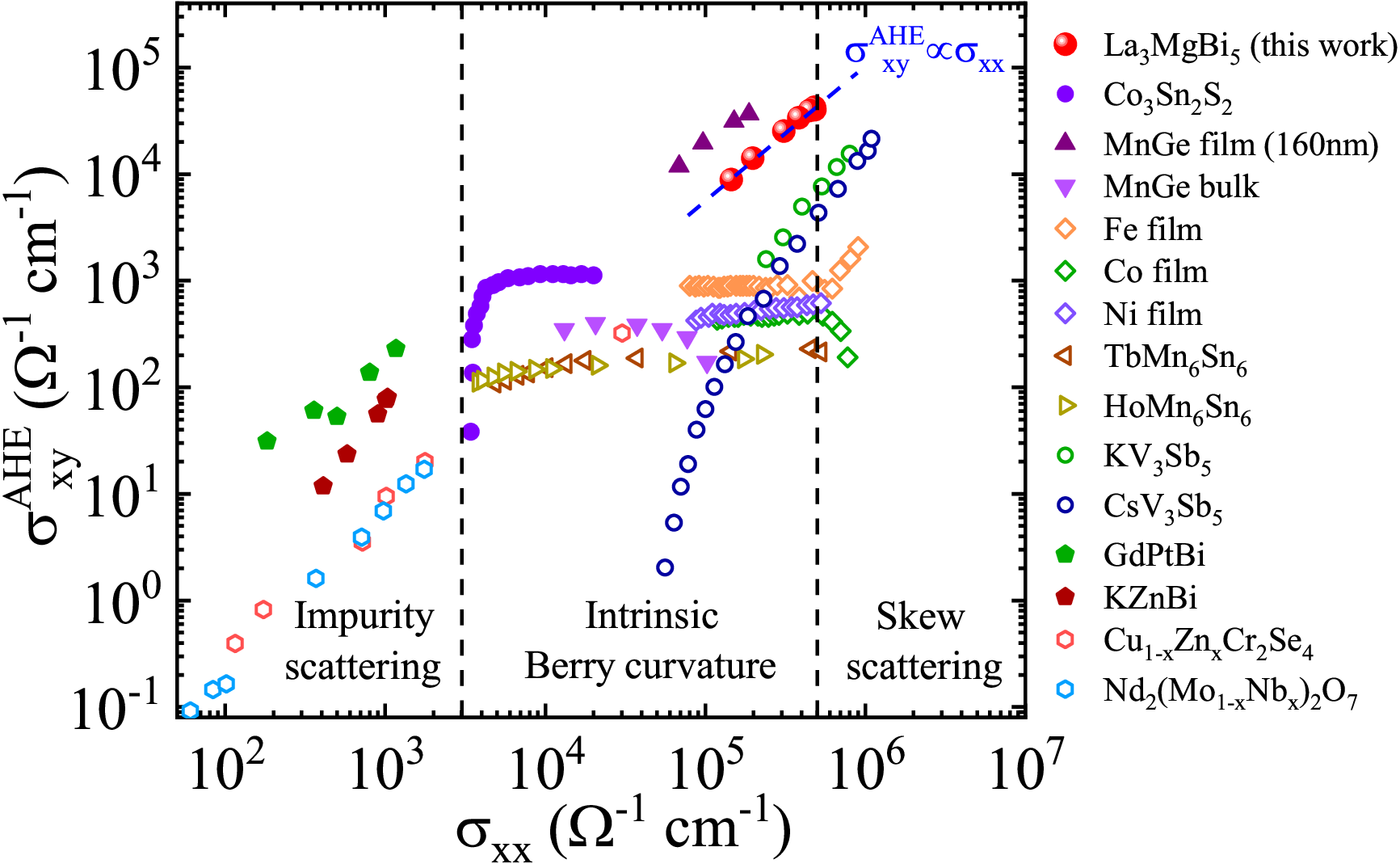,width=0.9\columnwidth}}
	\caption{\textbf{Map of AHE for various materials.} Logarithmic plot of $\sigma_{xy}^{\rm{AHE}}$ versus $\sigma_{xx}$. The anomalous Hall conductivity $\sigma_{xy}^{\rm{AHE}}$ of La$_3$MgBi$_5$ reaches 4.2356$\times$10$^4$ $\Omega^{-1}$ $cm^{-1}$ and follows the relationship of $\sigma_{xy}^{\rm{AHE}}$ $\propto$ $\sigma_{xx}$. The anomalous Hall conductivity was calculated by $\sigma_{xy}^{\rm{AHE}}$=$-\rho_{\rm{AHE}}^{\enspace 0}$/($\rho_{xx}^2$+($\rho_{\rm{AHE}}^{\enspace 0}$)$^2$).}
\end{figure*}

\begin{figure*}
	\centerline{\epsfig{figure=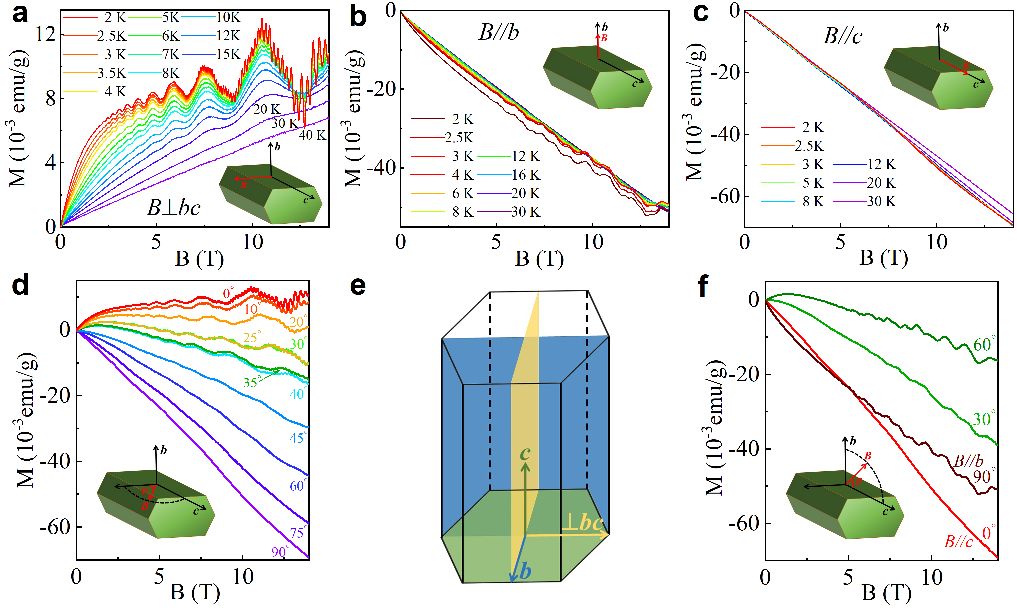,width=0.9\columnwidth}}
	\caption{\textbf{Axial diamagnetism in La$_3$MgBi$_5$ (Sample 3).} \textbf{a}-\textbf{c}, Magnetic field dependence of magnetization in three directions at various temperatures. The field is applied in the directions of $B \perp bc$, $B \parallel b$ and $B \parallel c$, respectively. \textbf{d}, Magnetic field dependent magnetization at different angles $\psi$. The direction of the field is rotated from $B \perp bc$ to $B \parallel c$. The temperature is fixed at 2 K. \textbf{e}, A schematic diagram of the measurement directions and corresponding vertical planes. \textbf{f}, Magnetic field dependent magnetization at different angles $\varphi$. The direction of the field is rotated from $B \parallel c$ to $B \parallel b$. The temperature is fixed at 2 K.}
\end{figure*}

\end{document}


\title{Supplementary Information for: Extremely large anomalous Hall conductivity and unusual axial diamagnetism in a quasi-1D Dirac material La$_3$MgBi$_5$}


\author{Zhe-Kai Yi,$^{1}$ Peng-Jie Guo,$^{2,\ddag}$ Hui Liang,$^{1}$ Yi-Ran Li,$^{1}$ Ping Su,$^{1}$ Na Li,$^{1}$ Ying Zhou,$^{1}$ Dan-Dan Wu,$^{1}$ Yan Sun,$^{1}$ Xiao-Yu Yue,$^{3}$ Qiu-Ju Li,$^{4}$ Shou-Guo Wang,$^{1}$ Xue-Feng Sun$^{1,5,\S}$ and Yi-Yan Wang$^{1,*}$}

	\maketitle
	
	\begin{affiliations}
		\item Anhui Key Laboratory of Magnetic Functional Materials and Devices, Institute of Physical Science and Information Technology, Anhui
            University, Hefei, Anhui 230601, China
        \item Department of Physics and Beijing Key Laboratory of Opto-electronic Functional Materials \& Micro-nano Devices, Renmin University of China, Beijing 100872, China
        \item School of Optical and Electronic Information, Suzhou City University, Suzhou, Jiangsu 215104, China
        \item School of Physics and Optoelectronic Engineering, Anhui University, Hefei, Anhui 230601, China
        \item Collaborative Innovation Center of Advanced Microstructures, Nanjing University, Nanjing, Jiangsu 210093, China
	\end{affiliations}
	
	\leftline{$^\ddag$Corresponding authors: guopengjie@ruc.edu.cn}
    \leftline{$^\S$Corresponding authors: xfsun@ahu.edu.cn}
    \leftline{$^*$Corresponding authors: wyy@ahu.edu.cn}\vspace*{1cm}
	
\newpage

\begin{figure*}
	\centerline{\epsfig{figure=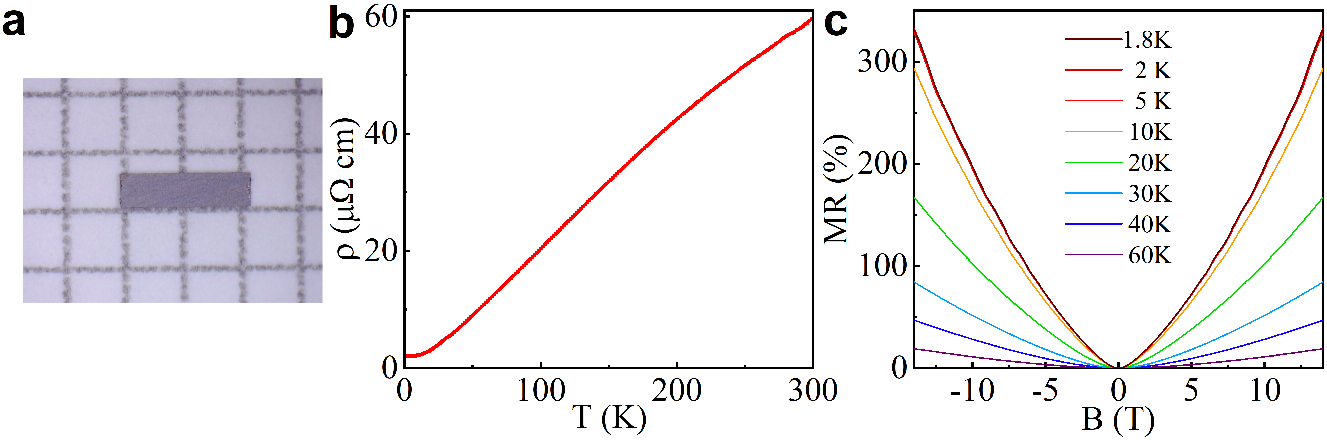,width=0.9\columnwidth}}
	\caption{\textbf{Resistivity and magnetoresistance of La$_3$MgBi$_5$ (Sample 4-b).} \textbf{a}, Sample 4 used in the measurements of AHE. Length = 2.22 mm, width = 0.58 mm. For Sample 4-a, the thickness is 0.09 mm. After the measurements of Sample 4-a, the sample was further processed to a thickness of 0.06 mm and marked as Sample 4-b. \textbf{b}, Temperature dependence of the resistivity. \textbf{c}, Magnetic field dependent MR at various temperatures. Weak Shubnikov-de Haas oscillation can be observed at low temperature and high field.}
\end{figure*}

\begin{figure*}
	\centerline{\epsfig{figure=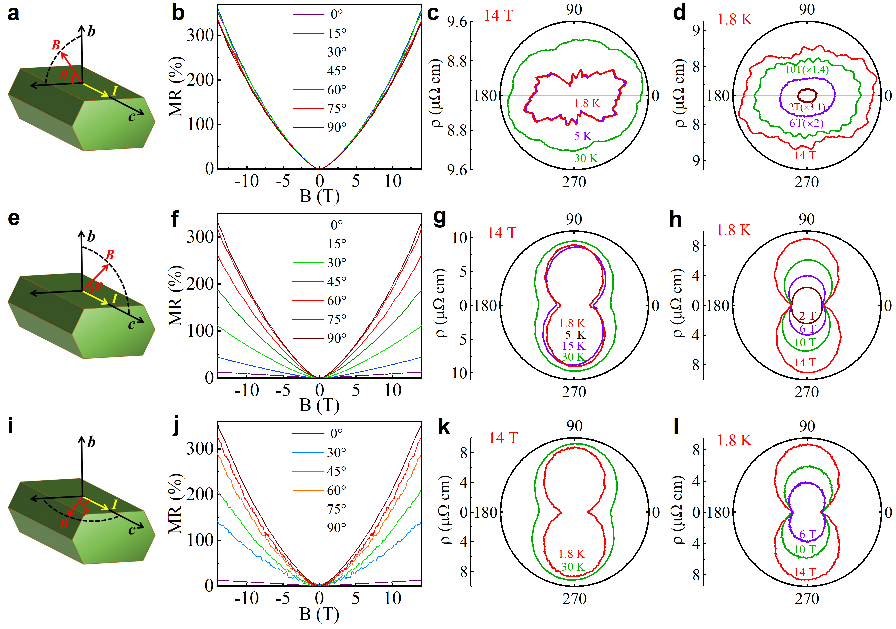,width=0.9\columnwidth}}
	\caption{\textbf{Angle dependence of the magnetoresistance.} \textbf{a}, \textbf{e} and \textbf{i}, Schematic of three configurations of the measurements. \textbf{b}, \textbf{f} and \textbf{j}, Magnetic field dependent MR at different angles corresponding to the configurations in \textbf{a}, \textbf{e} and \textbf{i}. The temperature is fixed at 1.8 K. \textbf{c}, \textbf{g} and \textbf{k}, Polar plots of the angle dependent resistivity. The magnetic field is fixed at 14 T and the temperature changes from 1.8 to 30 K. \textbf{d}, \textbf{h} and \textbf{l}, Polar plots of the angle dependent resistivity. The temperature is fixed at 1.8 K and the magnetic field varies from 2 to 14 T.}
\end{figure*}

\begin{figure*}
	\centerline{\epsfig{figure=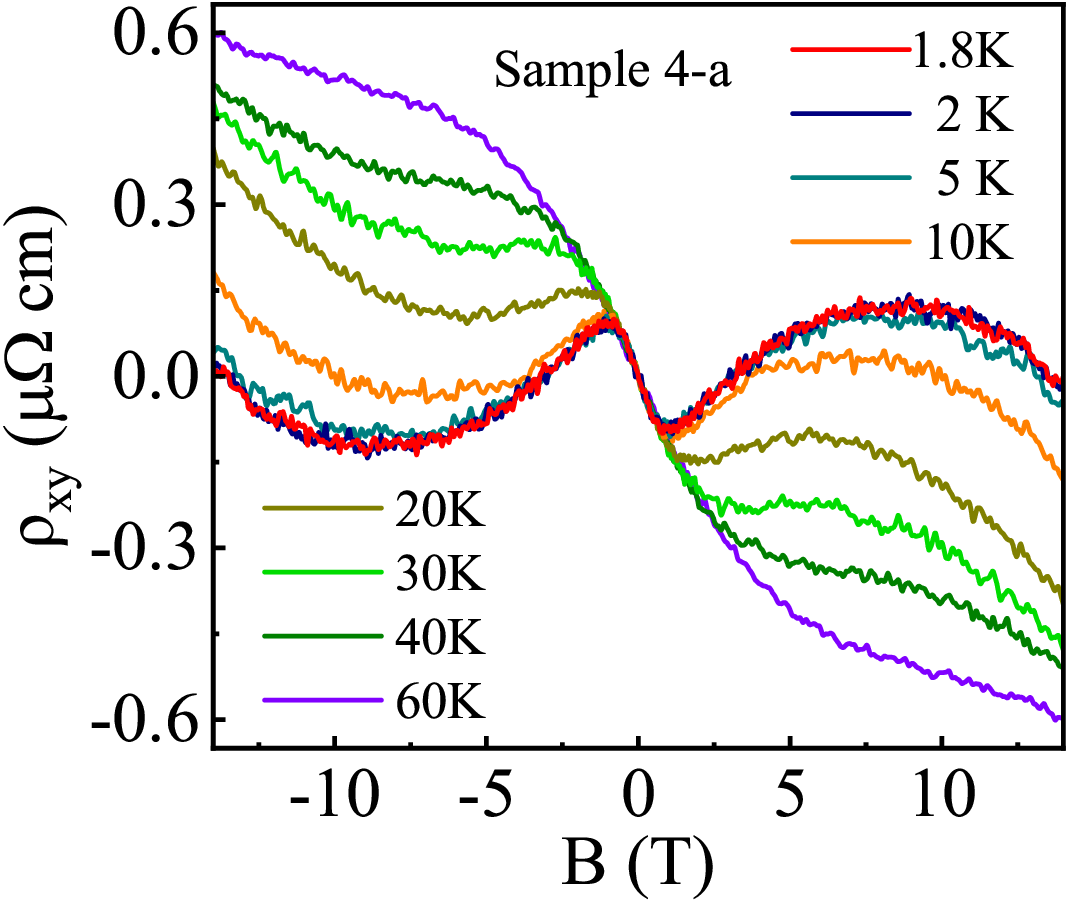,width=0.6\columnwidth}}
	\caption{\textbf{AHE in Sample 4-a.} The field dependent Hall resistivity of Sample 4-a at different temperatures. The thickness of Sample 4-a is 0.09 mm, which is thicker than Sample 4-b.}
\end{figure*}

\begin{figure*}
	\centerline{\epsfig{figure=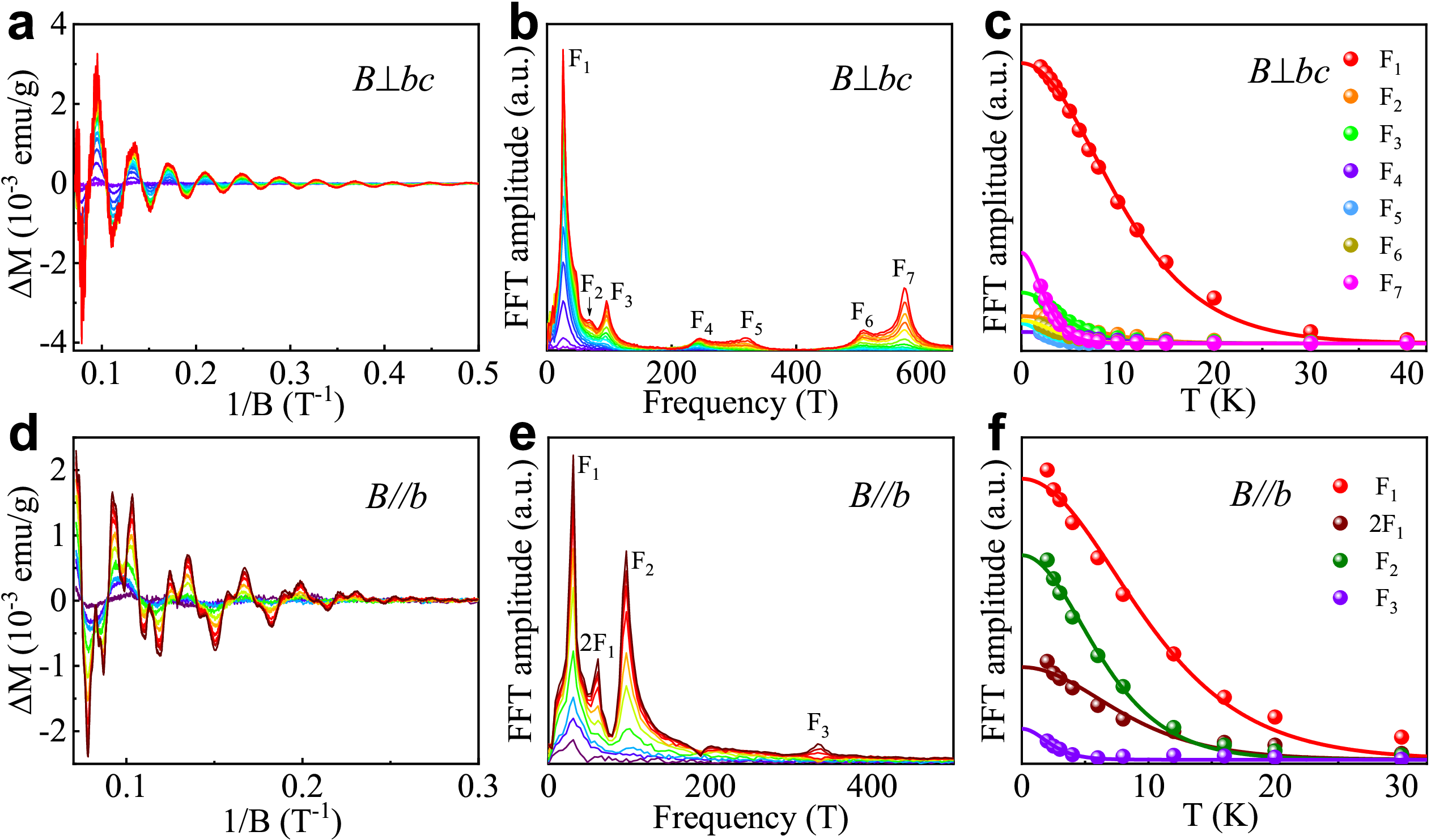,width=0.9\columnwidth}}
	\caption{\textbf{Analysis of the dHvA oscillations.} \textbf{a} and \textbf{d}, The oscillation amplitude extracted from Fig. 5a,b in the main text. The oscillation for $B \parallel c$ is too weak, making it difficult to extract the oscillation amplitude. \textbf{b} and \textbf{e}, The FFT spectra corresponding to \textbf{a} and \textbf{d}, respectively. \textbf{c} and \textbf{f}, Temperature dependent FFT amplitude of the frequencies. The solid lines are fittings using the thermal factor in Lifshitz-Kosevich formula. The derived parameters are listed in Table S1 below.}
\end{figure*}

\begin{figure*}
	\centerline{\epsfig{figure=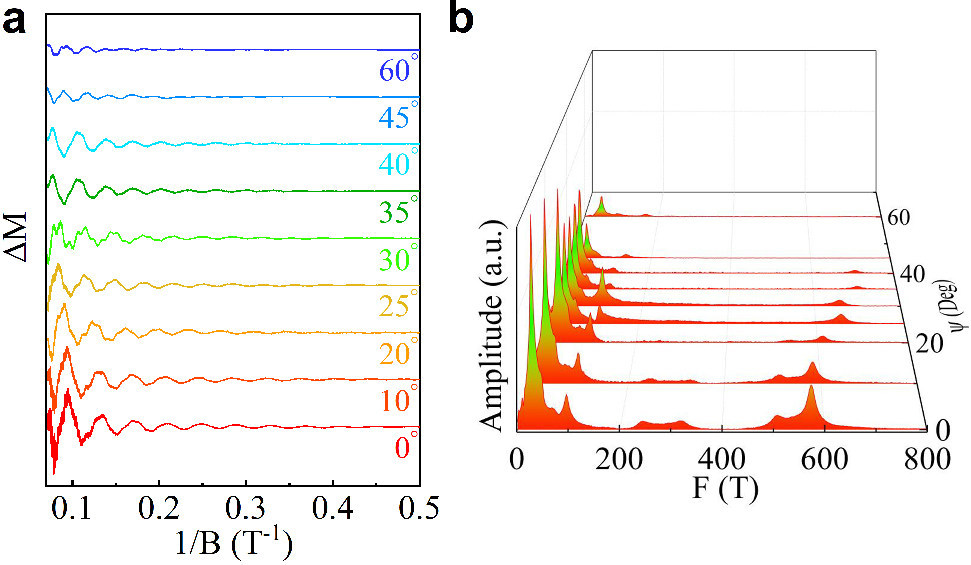,width=0.9\columnwidth}}
	\caption{\textbf{Analysis of the oscillations at different angles.} \textbf{a}, The oscillation amplitude extracted from Fig. 5d in the main text. \textbf{b}, A 3D plot of the FFT spectra corresponding to \textbf{a}.}
\end{figure*}

\begin{figure*}
	\centerline{\epsfig{figure=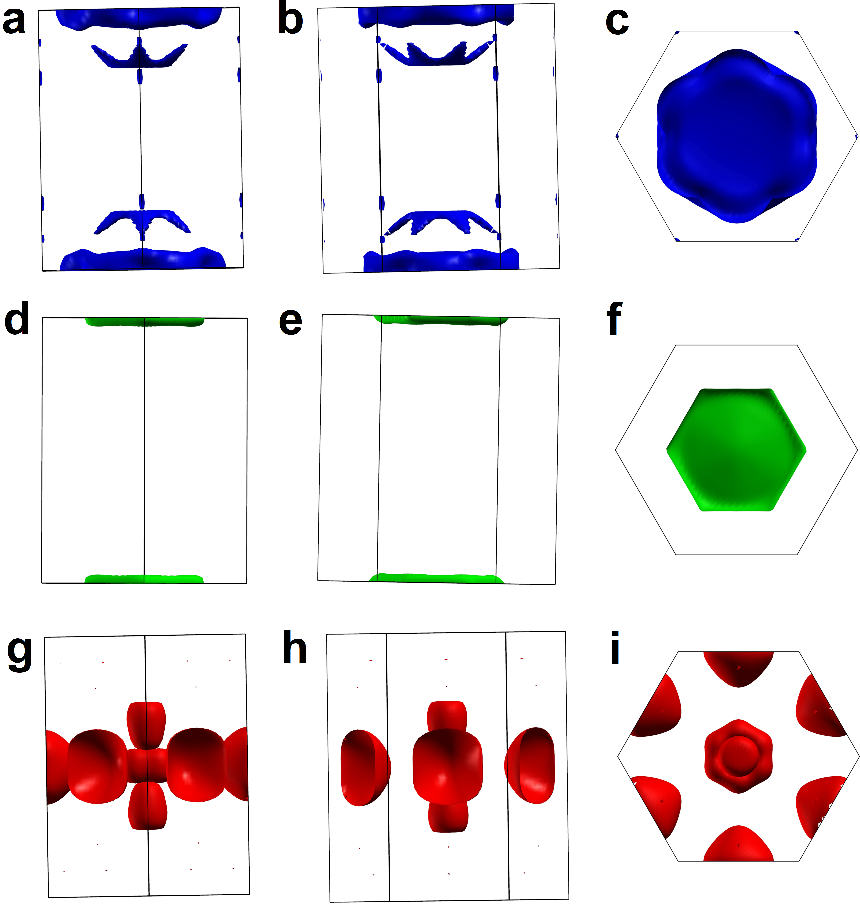,width=0.9\columnwidth}}
	\caption{\textbf{Fermi surfaces of La$_3$MgBi$_5$.} \textbf{a}-\textbf{c}, The electron-type Fermi surface corresponding to the blue band in Fig. 1e of the main text from different perspectives. \textbf{d}-\textbf{f}, Another electron-type Fermi surface corresponding to the green band from different perspectives. \textbf{g}-\textbf{i}, The hole-type Fermi surface corresponding to the red band from different perspectives.}
\end{figure*}

\begin{figure*}
	\centerline{\epsfig{figure=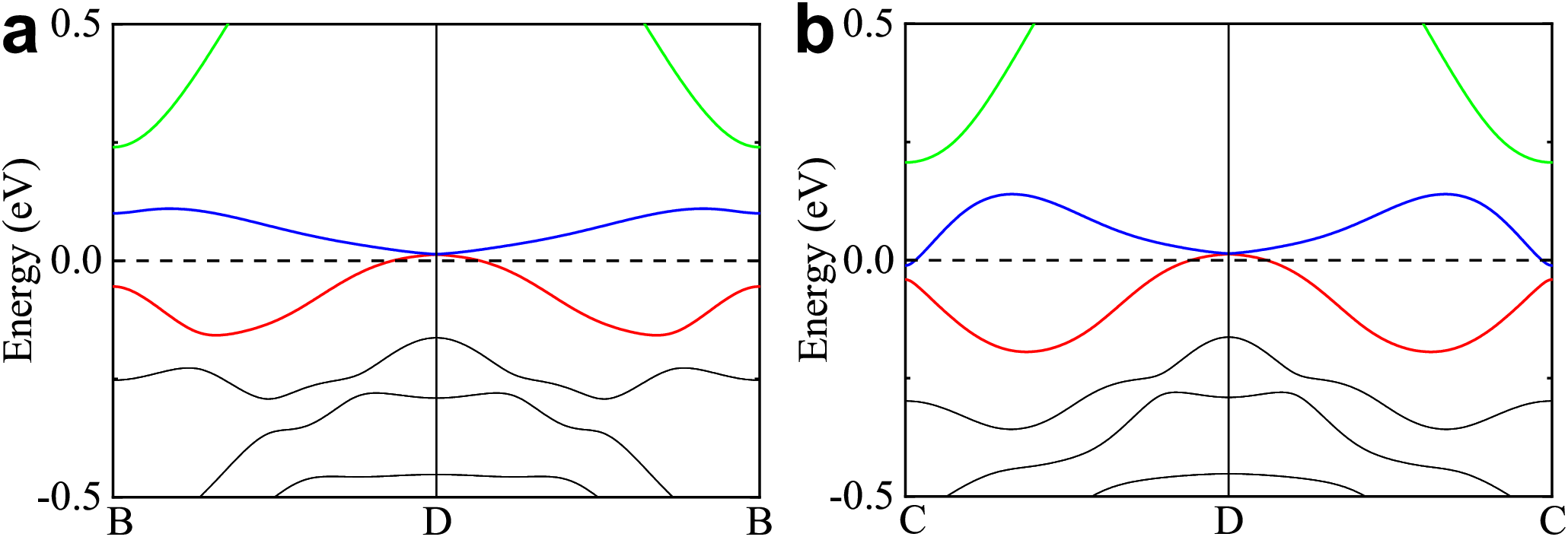,width=0.9\columnwidth}}
	\caption{\textbf{Band structure around the type-II Dirac point in the $\Gamma$-A direction.} \textbf{a}, Band structure along the B-D-B direction. \textbf{b}, Band structure along the C-D-C direction. D represents the position of the type-II Dirac point. The positions of B and C are marked in Fig. 1d.}
\end{figure*}

\begin{table*}
  \centering
  \caption{Parameters derived from the quantum oscillations. $F$, oscillation frequency; $A_F$, cross sectional area of Fermi surface normal to the field; $m^*$, effective mass.}
  \label{oscillations}
  \begin{tabular}{cccccccccccc}
    \hline\hline
          &   &    & \emph{F} (T) & \emph{A}$_F$ (10$^{-2}${\AA}$^{-2}$) & \emph{m}$^*$/\emph{m}$_e$ \\
    \hline
          & \multirow{7}{*}{$B \perp bc$} & \emph{F}$_1$ & 25.6 & 0.245 & 0.050 \\
          &    & \emph{F}$_2$ & 67.5 & 0.645 & 0.074 \\
          &    & \emph{F}$_3$ & 95.5 & 0.912 & 0.114 \\
          &    & \emph{F}$_4$ & 244.6 & 2.335 & 0.077 \\
          &    & \emph{F}$_5$ & 316.8 & 3.025 & 0.233 \\
          &    & \emph{F}$_6$ & 507.9 & 4.848 & 0.141 \\
          &    & \emph{F}$_7$ & 573.1 & 5.471 & 0.209 \\
          \cline{2-6}
          & \multirow{4}{*}{$B \parallel b$} & \emph{F}$_1$ & 30.6 & 0.292 & 0.081 \\
          &    & 2\emph{F}$_1$ & 61.2 & 0.584 & 0.099 \\
          &    & \emph{F}$_2$ & 96.2 & 0.918 & 0.126 \\
          &    & \emph{F}$_3$ & 332.4 & 3.173 & 0.340 \\
  \hline\hline
  \end{tabular}
\end{table*}